\documentclass{elsart}
\journal{New Astronomy}

\usepackage{natbib}

\usepackage{graphicx}
\usepackage{epsfig}

\usepackage{amssymb}

\begin{document}

\begin{frontmatter}



 \title{The 21 centimeter Background from the Cosmic Dark Ages: Minihalos 
and the Intergalactic Medium before Reionization}


\author{Kyungjin Ahn$^1$, Paul R. Shapiro$^1$, Marcelo A. Alvarez$^1$,}
\author{Ilian T. Iliev$^2$, Hugo Martel$^3$, and Dongsu Ryu$^4$}
\address{$^1$Department of Astronomy, 1 University Station, C1400,
  Austin, TX 78712, USA\\
$^2$Canadian Institute for Theoretical Astrophysics, University
  of Toronto, 60 St. George Street, Toronto, ON M5S 3H8, Canada\\
$^3$D\'epartement de physique, de g\'enie 
        physique et d'optique,          
        Universit\'e Laval, Qu\'ebec, QC G1K 7P4, Canada\\
$^4$Department of Astronomy and Space Science,
         Chungnam National University, Daejeon 305--764, Korea \\
E-mail: kjahn@astro.as.utexas.edu, shapiro@astro.as.utexas.edu,
marcelo@astro.as.utexas.edu, iliev@cita.utoronto.ca, 
hmartel@phy.ulaval.ca, ryu@canopus.cnu.ac.kr }

\begin{abstract}
The H atoms inside minihalos (i.e. halos with virial temperatures 
$T_{\rm vir} \le 10^{4} {\rm K}$, in the mass range roughly from
$10^{4} M_{\odot}$ to $10^{8} M_{\odot}$) during the cosmic dark ages
in a $\Lambda$CDM universe produce a redshifted background of
collisionally-pumped 21-cm line radiation which can be seen in
emission relative to the cosmic microwave background
(CMB). Previously, we used semi-analytical calculations of the 21-cm
signal from individual halos of different mass and redshift and the
evolving mass function of minihalos to predict the mean brightness
temperature of this 21-cm background and its angular fluctuations.
Here we use high-resolution cosmological N-body and hydrodynamic
simulations of  
structure formation at high redshift ($z\gtrsim 8$) to compute the
mean brightness temperature of this background from
both minihalos and the intergalactic medium (IGM) prior to the onset
of Ly$\alpha$ radiative pumping.
We find that 
the 21-cm signal  from gas in collapsed, virialized minihalos
dominates over that from
the diffuse shocked gas in the IGM.
\end{abstract}


\end{frontmatter}

\section{Introduction}
Neutral hydrogen atoms in the early universe can be detected in
absorption or emission against the cosmic microwave background (CMB)
at redshifted radio wavelength 21 cm, depending upon whether their
spin temperature $T_{\rm S}$ is less than or greater than that of the
CMB, respectively.
Minihalos which form during the
``dark ages'' have density and temperature high enough to appear in
emission (\citealt{2002ApJ...572L.123I,2003MNRAS.341...81I}). 
The intergalactic medium (IGM), on the other hand, appears
in either emission or absorption.
New radio telescopes are being designed to detect this 21 cm signal.

\citet[ISFM hereafter]{2002ApJ...572L.123I} showed that atomic collisions
inside minihalos are 
sufficient to decouple the spin temperature of the neutral
hydrogen from the CMB temperature. 
They predicted the mean and angular
fluctuations of the corresponding 21cm signal 
by a semi-analytical calculation based upon integrating the
individual minihalo contributions for different halo masses and
redshifts, over the evolving statistical distribution of minihalo
masses in the $\Lambda$CDM universe. The fluctuations, they found, are
substantial enough to be detectable with
future radio telescopes. \citet{2004ApJ...611..642F}, on the other hand,
have recently suggested that
the gas in the diffuse IGM (prior to the onset of Ly$\alpha$ radiative
pumping) 
is also capable of producing the 21cm
emission signal and that the IGM contribution to the mean signal
will dominate over that from gas inside minihalos.

In order to quantify these effects, 
we have computed the 21 cm signal both from minihalos and the
IGM at $z\gtrsim 8$, using  high-resolution cosmological N-body and
hydrodynamic simulations of structure formation. 
We use a flat, $\Lambda$CDM cosmology with matter 
density parameter $\Omega_{m}=0.27$, cosmological constant 
$\Omega_{\Lambda}=0.73$, baryon density $\Omega_b=0.043$, Hubble constant 
$H={\rm 70\, km\,s^{-1}Mpc^{-1}}$, $\sigma_{8h^{-1}}=0.9$ and the
Harrison-Zel'dovich primordial power spectrum. 

\section{THE CALCULATION}
\label{sec:calculation}

\subsection{Basics of 21 cm Radiation Background from the Dark Ages}
\label{sec:basic}
Foreground emission or absorption by neutral hydrogen atoms at
redshift $z$ is seen against the CMB at redshifted wavelength, $21(1+z)\,
{\rm cm}$. The spin temperature ($T_{\rm S}$) of a hydrogen atom
determines whether the 
signal is in emission or absorption. Emission occurs when $T_{\rm
  S} > T_{\rm CMB}$, while absorption occurs when $T_{\rm
  S} < T_{\rm CMB}$.
$T_{\rm S}$ can deviate from $T_{\rm CMB}$ in various ways. A
hydrogen atom can 1) absorb a 21 cm photon from CMB (CMB pumping), 2)
collide with another atom (collisional pumping) and 3) absorb a
Ly$\alpha$ photon to make a Ly$\alpha$ transition, then decay to one of
hyperfine 21 cm levels (Ly$\alpha$ pumping).  

During the dark ages, when there is no Ly$\alpha$ pumping, the mean 21 cm
signal against the CMB will be zero at $z\gtrsim 150$, then in
absorption at $20\lesssim z\lesssim 150$ mostly due to unperturbed
IGM, and finally in emission at $z\lesssim 20$ due primarily to minihalos. We
restrict ourselves to regions such that the Ly$\alpha$ pumping is
negligible even after sources turn on at $z\lesssim 20$. In other
words, we only consider collisional pumping in minihalos and the IGM
at $8\lesssim z\lesssim 100$.

\subsection{Semi-analytic calculations}
\label{sub:Analytic-calculation}
\begin{figure}[!ht]
\centerline{\psfig{file=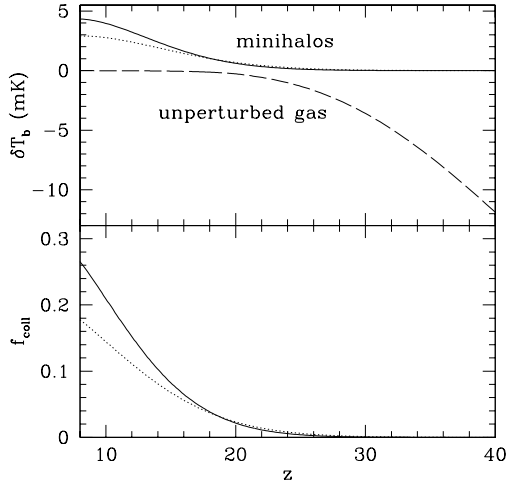,width=2.7in}}
\caption{\label{fig-anal}Analytical prediction for the mean 21 cm 
differential brightness temperature due to collisionally-decoupled 
minihalos and an unperturbed IGM. Shown are the results based on the 
Press-Schechter (solid) and the Sheth-Tormen (dotted) mass functions 
for halos and the contribution from the IGM gas with cosmic mean 
density and temperature (dashed). In the bottom panel we show the 
minihalo collapsed fraction, again based on the Press-Schechter (solid) and 
the Sheth-Tormen (dotted) mass functions.}
\end{figure}

Here we briefly summarize the semi-analytical calculation of the 21 cm
signal from minihalos by ISFM and from the unperturbed IGM. For minihalos, the
differential brightness temperature,
\begin{equation}
\label{eq-dT}
\delta T_{b}=\frac{T_{\rm S}-T_{{\rm CMB}}(z)}{1+z}(1-e^{-\tau}),
\end{equation}
is averaged over individual minihalos to give the mean differential
antenna temperature $\overline{\delta T}_{b}$ (for detail, see
ISFM). On the other hand, the unperturbed IGM of the universe has a
kinetic temperature smaller than the CMB temperature at $z\lesssim
100$, resulting in an absorption signal until  collisional pumping
becomes negligible at $z\simeq 20$ (Fig. [\ref{fig-anal}]) .

\subsection{Numerical Simulations}
\label{sec:Numerical-Simulations}

We have run a high resolution cosmological N-body and gasdynamic
simulation to derive the effects of gravitational collapse and 
hydrodynamics on the predicted 21 cm signal from the high redshift
universe. Our computational box has a comoving size of 
$0.7\, {\rm Mpc}$, 
with $1024^3$ cells and $512^3$ dark matter particles,
which is optimal for adequately resolving both the
minihalos and  small-scale structure-formation shocks. We have used the code
described in \citet{1993ApJ...414....1R}, which uses the particle-mesh
(PM) scheme for calculating the gravity evolution and an Eulerian
total variation diminishing (TVD) scheme for hydrodynamics. 

In order to calculate the minihalo and the IGM contributions to the
total differential 
brightness temperature, one needs first
to identify the halos in the simulation volume. We identified the halos 
using a friends-of-friends (FOF) algorithm
(\citealt{1985ApJ...292..371D}) with a 
linking length parameter of $b=0.25$. The rest of the signal is
considered to be from the IGM. 

\begin{figure}[!ht]
\centerline{\psfig{file=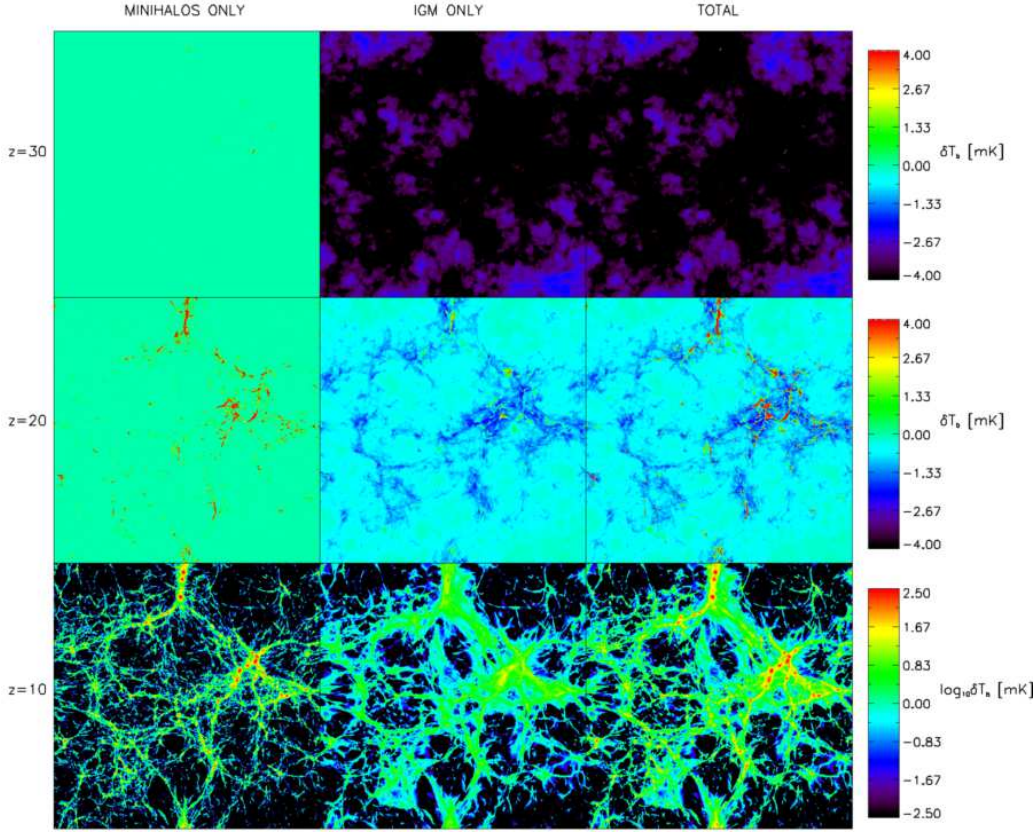,width=5.4in}}
\caption{
\label{maps}Map of the 21 cm signal obtained from our high
  resolution simulation. Rows, top to bottom, show redshifts $z$=30, 20,
  and 10. Columns, left to right, represent contributions from minihalos, the
  IGM and the total signal, respectively. Note that the scale is
  linear for the upper two 
  rows of images, but logarithmic for the bottom row. 
}
\label{fig:prim}
\end{figure}

\begin{figure}[!ht]
\centerline{\psfig{file=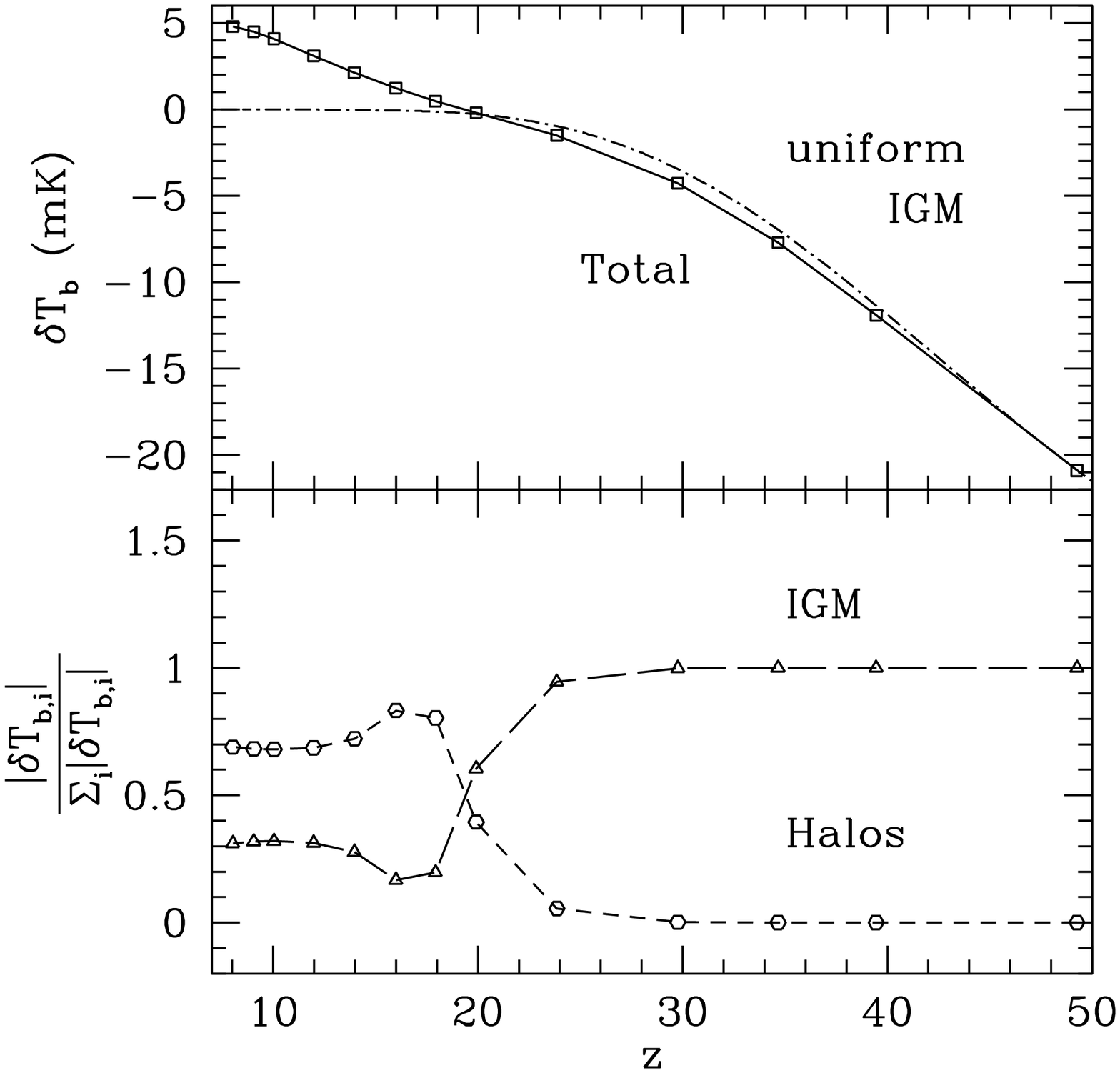,width=2.7in}\psfig{file=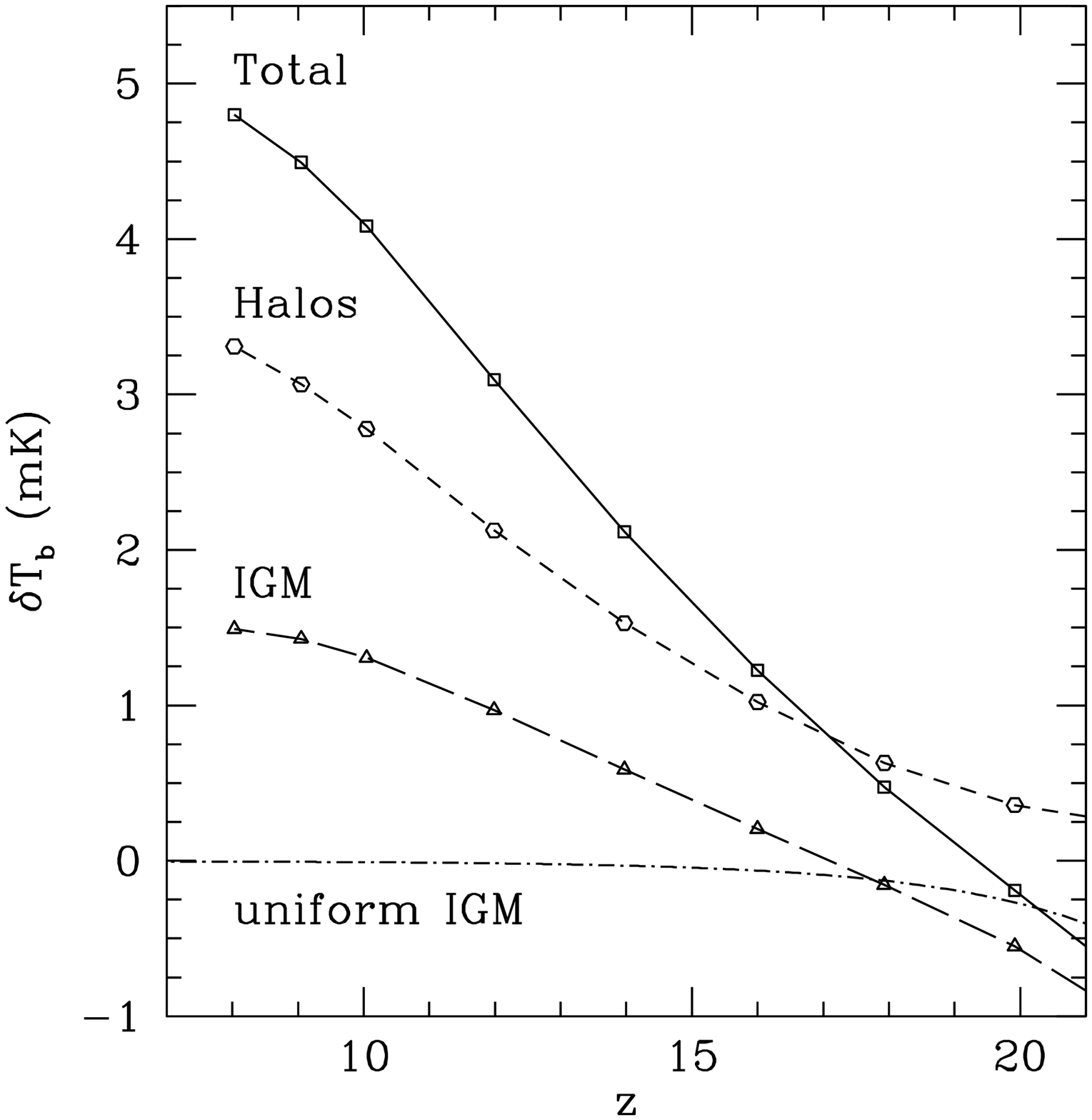,width=2.7in}}
\caption{\label{fig-0.25-numerical}21 cm mean brightness temperature evolution.
(a)(left) Evolution of the total 21-cm signal vs. redshift. 
All data points are directly calculated from our simulation box,
with the assumption that optical depth is negligible throughout the box.
(b)(right) $\delta T_b$ vs. redshift below $z=20$. Plotted are the
contributions from minihalos (labeled Halos, circles), the IGM (triangles)
and the total (squares).}
\end{figure}

In Figure \ref{maps}, we show projected maps for three different
redshifts. Each contribution from minihalos and the IGM is also
plotted. It clearly shows that the signal is in absorption due to the IGM at
$z\gtrsim 20$, and after crossing the moment when a mixture of emission
from minihalos and absorption from the IGM exists ($z\approx 20$),
the signal is predominantly in emission both from minihalos and the
IGM at $z\lesssim 20$. In Figure \ref{fig-0.25-numerical}, we show the
total signal as well as minihalos and the IGM contributions.

\subsection{Semi-Analytical Calculation of the Halo Contribution}
\label{sub:Improvements}

Our numerical simulations have sufficiently high resolution to find
all of the minihalos
in the computational box,  as well as the large-scale structure
formation shocks, but 
not to resolve the internal structure of the minihalos themselves. 
As ISFM have shown, in order to obtain the correct 21-cm signal from minihalos
one needs to do a full radiative transfer calculation through each
individual minihalo density profile since, unlike the IGM
gas, minihalos have a non-negligible optical depth at the 21-cm line.
Hence, we can refine our estimate of the minihalo contribution to
the total 21-cm signal by combining our numerical halo catalogues with the
semi-analytic calculation of individual minihalo contribution as found by ISFM. 

In their approach, the gas density of each minihalo is assumed to
follow the TIS 
profile, of \citet{2001MNRAS.325..468I}, the radiative
transfer calculation is performed for different impact parameters,
and then finally the face-averaged $\delta T_{b}$ is calculated 
(see ISFM for details), according to
\begin{equation}
\label{eq-dTb}
\overline{\delta T}_{b}=\frac{c(1+z)^{4}}{\nu_{0}H(z)}\int_{M_{{\rm
      min}}}^{M_{{\rm max}}}\Delta\nu_{{\rm eff}}\delta
T_{b,\,\nu_{0}}A\frac{dn}{dM}dM.
\end{equation}
The halo mass function, $dn/dM$, is provided here, not by the
analytical PS mass function as in ISFM, however, but by the halo catologue
we construct from the simulation. Each individual halo contribution, 
$\Delta\nu_{{\rm eff}}\delta T_{b,\,\nu_{0}}A$, depends on its mass and
redshift of formation (ISFM). Once we calculate
$\Delta\nu_{{\rm eff}}\delta T_{b,\,\nu_{0}}A$, we then obtain the
halo contribution using equation~(\ref{eq-dTb}).

\begin{figure}[!ht]
\centerline{\psfig{file=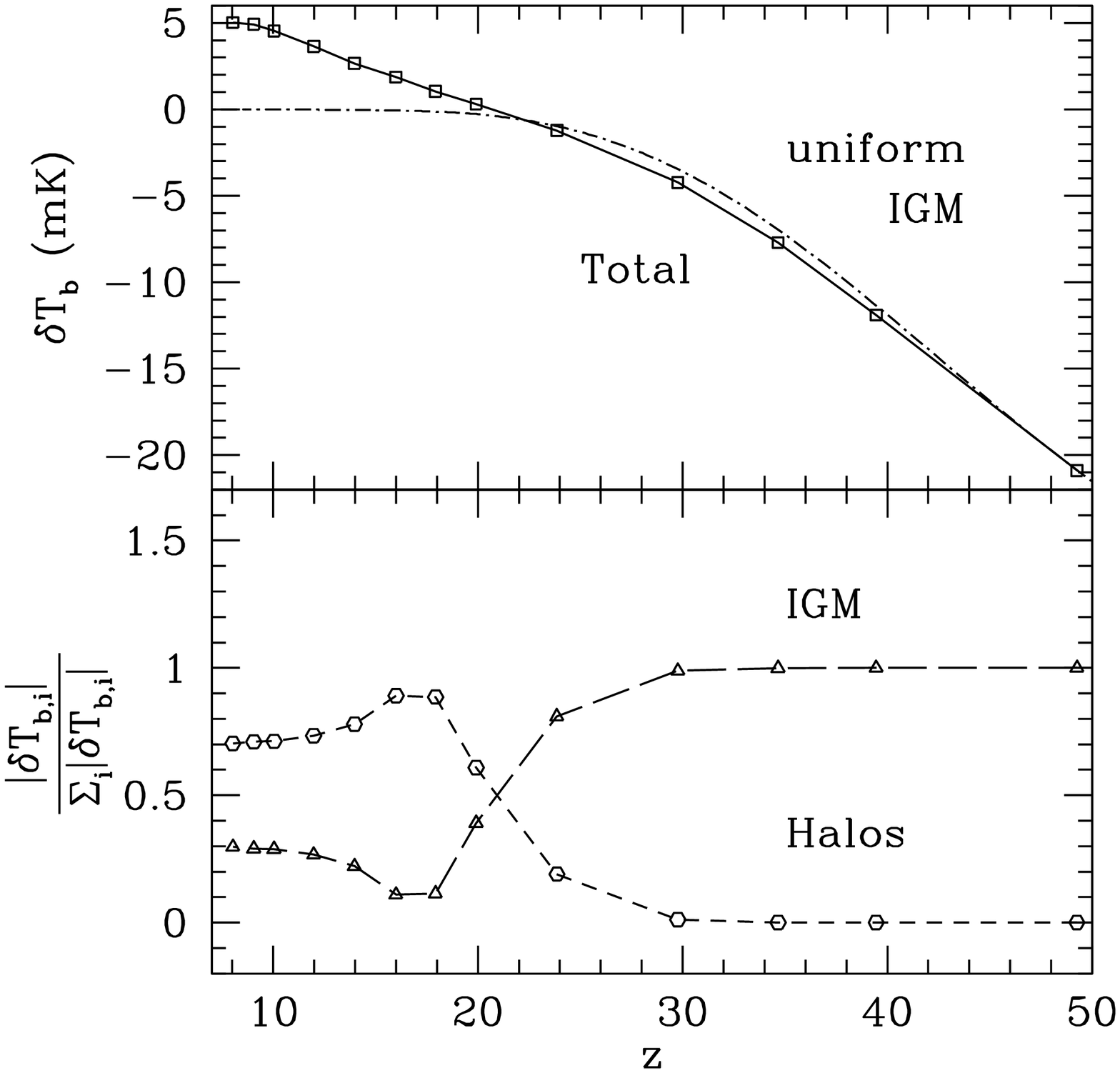,width=2.7in}\psfig{file=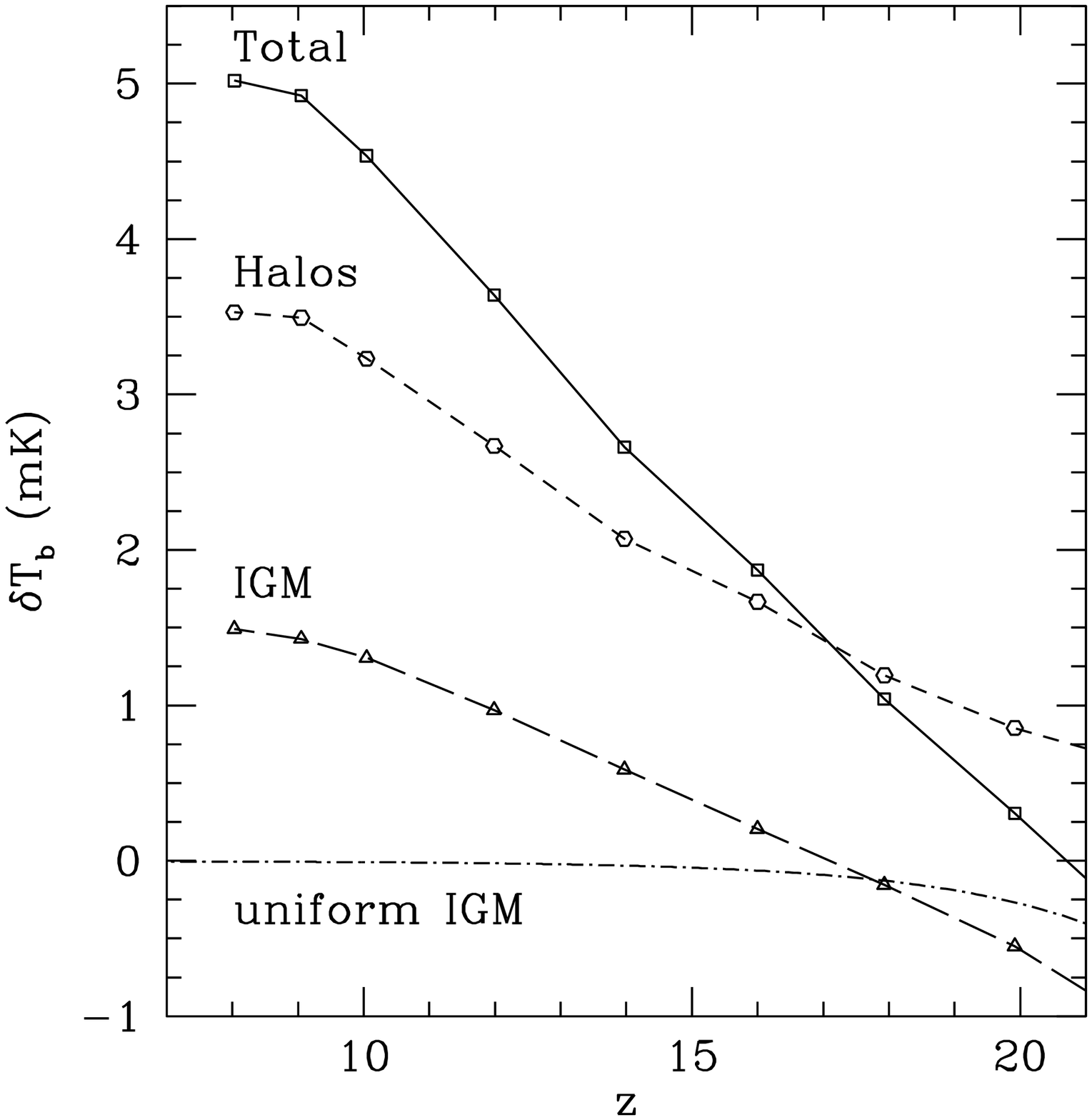,width=2.7in}}
\caption{\label{fig-0.25-improved} Semi-analytical minihalo signal vs. IGM
signal. The 21 cm flux from each halo in the simulation is found by modeling
the halos as described in \S~\ref{sub:Improvements}, to estimate the halo 21
cm signal more accurately. Same notation as in
Figure~\ref{fig-0.25-numerical}. The 21 cm minihalo emission increases
compared to the raw minihalo signal in figure~\ref{fig-0.25-numerical}. The
IGM signal remains the same. 
}
\end{figure}

\section{Minihalos vs. the IGM}
\label{sec:halo_vs_IGM}
We find that the emission signal at $z\lesssim 20$ is dominated by
minihalos. For $8 \lesssim z \lesssim 14$, about $70-80 \%$ of the
total signal is contributed by minihalos. The minihalo contribution peaks
at $z\simeq 16 - 18$, because the IGM is in both emission and
absorption. At $z\simeq 20$, the IGM is mostly in absorption, with an
absolute amplitude which is comparable to the minihalo emission signal. Only
when the collapsed fraction of baryons into minihalos is negligible
($z\gtrsim 20$), does the IGM make the dominant contribution to the total
signal, which is in absorption. Note that the total signal at $z\gtrsim
20$ is slightly stronger than the analytical value for the unperturbed
IGM, because clumping of the IGM makes the collisional coupling of the
spin temperature to the kinetic temperature slightly stronger.

\begin{figure}[!ht]
\centerline{\psfig{file=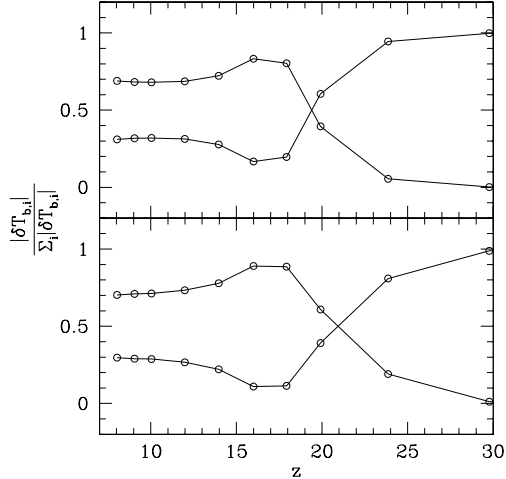,width=2.7in}}
\caption{\label{fig-conv_rel} Fractional contributions of minihalos and diffuse
  IGM gas to the total 21-cm background. The top panel shows the results
  obtained directly from simulations (C1: triangle, long-dashed; C2:
square, short-dashed; C3: pentagon, dotted; C4: circle, solid). 
The bottom panel shows the results which were semi-analytically
refined (\S~\ref{sub:Improvements}; point- and line-types follow those
of the top panel). }
\end{figure}

\section{Conclusions}
\label{sec:Conclusion}

We have performed a cosmological N-body and hydrodynamic simulation of the
evolution of dark matter and baryonic gas in the $\Lambda$CDM universe
at high redshifts ($8 \lesssim z \lesssim 100$). With
the assumption that radiative feedback effects from the first light sources
are negligible, we calculated the 21 cm mean differential brightness
temperature signal. The mean global signal is in absorption against the CMB
above $z\simeq20$ and in overall emission below $z\simeq18$. At
$z \gtrsim 20$, the
density fluctuations of the IGM gas are largely in the linear regime,
and their absorption 
signal is well approximated by the one that results from assuming uniform gas
at the mean adiabatically-cooled IGM temperature. At $z \lesssim 20$, nonlinear
structures become common, both minihalos and clumpy, warm, mildly nonlinear
IGM, resulting in an overall emission at 21 cm with differential brightness
temperature of order a few mK.

By identifying the halos in our simulations, we were able to separate and
compare the relative contributions of the halos and the IGM gas to the total
signal. We find that the emission from minihalos 
dominates over that from the IGM outside minihalos, for $z \lesssim
20$. The minihalos contribute $70-80\,\%$ of
the total emission signal at $z<16$, exceeding $\gtrsim 85\,\%$ at
$z\simeq 16-18$, and balancing the absorption by the IGM gas at $z\approx20$.
In contrast, the absorption by cold IGM gas dominates the total signal for
$z>20$, with a signal close to that of an unperturbed IGM.


\end{document}